\documentclass[a4paper]{revtex4}
\pdfoutput=1
\usepackage{graphicx}
\usepackage{fancyhdr}
\usepackage{amsmath}
\usepackage{color}

\usepackage{calc}
\usepackage{rotating}
\usepackage[english]{babel}
\usepackage{subfig}
\usepackage{float}
\usepackage{amssymb}
\usepackage{amsthm}
\usepackage{latexsym}
\usepackage{dcolumn}
\usepackage{multirow}
\usepackage{hyperref}
\usepackage{mciteplus}
\usepackage{enumitem}
\usepackage{slashed}

\newcommand*{\hpm}{\ensuremath{H^{\pm}}}
\newcommand*{\gev}{\ensuremath{\,\mathrm{GeV}}}
\newcommand*{\tev}{\ensuremath{\,\mathrm{TeV}}}
\newcommand*{\hobs}{\ensuremath{H_\text{obs}}}
\newcommand*{\hsm}{\ensuremath{H_\text{SM}}}
\newcommand*{\mhpm}{\ensuremath{m_{\hpm}}}
\newcommand*{\met}{\ensuremath{\slashed E_T}}

\begin{document}

\title{$H^\pm$ in the $W^\pm h$ channel at the LHC Run 2}

%

\author{R. Enberg$^a$}
\author{W. Klemm$^a$}
\author{S. Moretti$^b$}
\author{S. Munir$^{a,c,*}$}
\author{G. Wouda$^a$}

\affiliation{$^a$Department of Physics and Astronomy, Uppsala University, Box 516, SE-751 20 Uppsala, Sweden}
\affiliation{$^b$School of Physics \& Astronomy, University of Southampton, Southampton SO17 1BJ, UK}
\affiliation{$^c$Asia Pacific Center for Theoretical Physics, San 31, Hyoja-dong, Nam-gu, Pohang 790-784, Republic of Korea}

\begin{abstract}
We analyse the discovery prospects of the charged Higgs boson, $H^\pm$, via its decay in the $W^\pm h$ channel in the minimal supersymmetric Standard Model (MSSM) as well as several 2-Higgs Doublet Models (2HDMs). $h$, the lightest scalar Higgs boson in these models, is  identified with the recently discovered $\sim 125$\,GeV state, \hobs, at the Large Hadron Collider (LHC). 
We find that, while it provides an important input in the kinematic selection of signal events, the measured \hobs\ mass renders this channel inaccessible in the MSSM. In the 2HDMs though, through a dedicated signal-to-background analysis for the $pp\to t(\bar{b})H^-\to \ell^\pm\nu_\ell jj bb\bar{b}(\bar{b})$+h.c process, we establish that some parameter space regions will be testable at the LHC with $\sqrt{s}=14$\,TeV. 
\end{abstract}

\maketitle

\thispagestyle{fancy}


\section{Introduction}

The MSSM and the 2HDMs contain two scalar electroweak (EW) doublets, $\Phi_1$ and $\Phi_2$, which results in a total of five physical states, two scalars, $h,\,H$, one pseudoscalar, $A$, and a charged pair, $\hpm$, upon EW symmetry breaking. The mass of the $H^\pm$ is a free 
parameter in these models. When the $H^\pm$ is heavier than the top quark, its main
production process at the LHC is $pp\rightarrow tH^-\bar{b}$ (by which we actually 
imply $pp\rightarrow tH^-\bar{b}$ + $pp\rightarrow \bar{t}H^+b$; here and onwards,
 we do not distinguish between fermions and anti-fermions when their identity is 
unspecified and/or can be inferred from the context), and it decays
dominantly into $tb$. However, this decay channel is very difficult to
probe when the parameter $\tan\beta$
(the ratio of the vacuum expectation values of the two Higgs fields)
has a value less than $\sim 3$, owing to the large reducible and irreducible
backgrounds. For such $\tan\beta$, earlier
studies~\cite{Moretti:2000yg} have shown that the decay
channel $H^\pm \to W^\pm h$ could prove crucial in the MSSM. 

Since the discovery of \hobs\ at the LHC~\cite{Aad:2012tfa,Chatrchyan:2012ufa}, its mass measurement, $m_{\hobs}\approx 125$\gev, provides an additional input for collider analyses
 as well as an important constraint on physics beyond the Standard Model (SM). 
We therefore revisit the $H^\pm \to W^\pm h$ decay channel, assuming $h$ to be consistent with the \hobs, in the MSSM, the 2HDM of Types I and II (inferences drawn for which can largely be extended, respectively, to Types III and IV also), and the aligned 2HDM (A2HDM). In all these models, instead of the $h$, the $H$ can also play the role of the \hobs, but we do not consider this scenario here. We estimate the signal strength of the considered process in the considered models' parameter space regions satisfying some important experimental constraints. We then compare these estimates to the expected sensitivity obtained from a signal-to-background analysis for the LHC with $\sqrt{s}=14$\,TeV.  

\section{Models and parameters}

As noted above, the production process of interest here is $pp\rightarrow tH^-\bar{b}$. When calculating the cross section for this process, the two subprocesses, $gg\rightarrow tH^-\bar{b}$ and $gb\rightarrow tH^-$, cannot be added na\"{i}vely, as that results in a double counting between the two contributions. This is because the $gg$ amplitude can be seen as a tree-level contribution to the next-to-leading order amplitude that includes a virtual $b$-quark, with the $gb$ process making the leading order amplitude. The square of the amplitude for the $gb \rightarrow tH^-$ process, with spin/colour summed/averaged, is given by~\cite{Moretti:2002eu}
\begin{eqnarray}
\label{eq:sigma}
\overline{|\mathcal{M}|^2}
=\frac{g^2_{q\hpm}}{2m_W^2} \, \frac{g_s^2 g_2^2}{4N_c}|V_{tb}|^2 \,  
\frac{(u-\mhpm^2)^2}{s(m_t^2-t)}\left[1+2\frac{\mhpm^2-m_t^2}{u-\mhpm^2}\left(1+\frac{m_t^2}{t-m_t^2}+\frac{\mhpm^2}{u-\mhpm^2}\right)\right],
\end{eqnarray}
where $g_s$ is the $SU(3)_C$ and $g_2$ the $SU(2)_L$ gauge couplings, $N_C=3$ is
the number of colours and $V_{tb}$ is the relevant CKM matrix element. The model-dependence of this process is embedded in the coupling $g_{q\hpm}$, along with the predicted $\hpm$ mass. The $\hpm \rightarrow W^\pm h$ decay channel, which is proportional to the coupling 
\begin{equation}
\label{eq:hpmhwcpl}
g_{h H^+ W^-} = \frac{g_2}{2}(\cos\beta S_{12} - \sin\beta S_{11}) \,,
\end{equation}
where $S_{11}$ and $S_{12}$ are the elements of the scalar Higgs mixing matrix, has a strong dependence on $\tan\beta$, similarly to the production process, as we will see below.

In the MSSM the squared $\hpm$-quark coupling noted above is given by
$g^2_{q\hpm} = m_b^2\tan^2\beta + m_t^2\cot^2\beta$,
where $m_b$ and $m_t$ are the masses of the bottom and top quarks, respectively.
In addition to the most relevant parameters here, $\tan\beta$ and $m_{\hpm}$, some other free parameters of the MSSM are also crucial, particularly for obtaining the desired mass of $h$. These include the Higgs-higgsino mass parameter, $\mu$, the soft masses and trilinear couplings of the sfermions and the soft gaugino masses. (These supersymmetric (SUSY) parameters are taken into account here under certain unification assumptions; see~\cite{Enberg:2014pua} for further details about them and their scanned ranges.)
 
In the 2HDMs, the physical masses of the five Higgs bosons as well as the fermion Yukawa couplings are all free parameters. However, in order to avoid large tree-level flavour-changing neutral currents (FCNCs), a $Z_2$-symmetry is generally introduced, so that each type of fermion only couples to a single doublet~\cite{Glashow:1976nt,Paschos:1976ay}.  In 2HDM Type I (2HDM-I), the fermions only couple to $\Phi_2$ and in Type II (2HDM-II), the down-type quarks and leptons couple to $\Phi_1$ and up-type quarks to $\Phi_2$. The Yukawa couplings are then determined entirely by the parameter $\tan\beta$. In the CP-conserving limit assumed here, $\tan\beta$, $|\sin(\beta-\alpha)|$ and $m_{12}^2$ are then the only additional free parameters. Another mechanism for controlling FCNCs is to require that the two Higgs doublets have Yukawa matrices which are proportional to one another, or aligned. Here we consider the case where all fermions couple to both $\Phi_1$ and $\Phi_2$ with aligned couplings, known as the A2HDM. In this model, $|\sin\alpha|$, $\lambda_2$, $\lambda_3$, $|\lambda_7|$ and $|\beta^{U,D,L}|$ are the only free parameters besides the Higgs boson masses. (See~\cite{Pich:2009sp} for a complete description of these parameters.) The coupling
$g^2_{q\hpm}$ in each of these 2HDMs is given in table~\ref{table:g2qHm}.
\begin{table}[ht!]
\vspace*{-0.4cm}
\begin{center}\caption{\label{table:g2qHm}
 The expressions for $g^2_{q\hpm}$ in the 2HDMs considered in our analysis.}
\begin{tabular}{|c|c|c|c|}
\hline
 & 2HDM-I & 2HDM-II & A2HDM \\
\hline
 $g^2_{q\hpm}$ & $m_b^2\cot^2\beta + m_t^2\cot^2\beta $ &   $ m_b^2\tan^2\beta + m_t^2\cot^2\beta $ & $ m_b^2\tan^2\beta^D + m_t^2\tan^2\beta^U $ \\
\hline
\end{tabular}
\end{center}
\end{table}

\vspace*{-0.2cm} 
We performed scans of the above mentioned parameter space of each model to search for their regions with a strong signal. The particle mass spectra and decay branching ratios (BRs) in the MSSM and the 2HDMs were generated using the programs SUSY-HIT-v1.3~\cite{Djouadi:2006bz} and 2HDMC~\cite{Eriksson:2009ws}, respectively. In these scans we required $123\leq m_h \leq 127 \gev$ and $200\leq \mhpm\leq 500\gev$, and additionally in the case of the 2HDMs, $135\leq m_H\leq 500\gev$ and $m_A=\mhpm$. We allowed a slightly extended range,  $1.5\leq\tan\beta\leq 6$.
For the MSSM and AHDM, the $b$-physics observables BR$(\bar{B}\to X_s\gamma)$, BR$(B_u\to \tau\nu)$, and BR$(B_s\to \mu^+\mu^-)$ were calculated using SuperIso-v3.4, and were required to lie within the 95\% confidence level limits suggested in the manual of the program~\cite{superiso}. For the $Z_2$-symmetric 2HDMs, the relevant input parameters were simply chosen so as to satisfy these constraints, following~\cite{Mahmoudi:2009zx}. Finally, each scanned point was tested with HiggsBounds-v4.1.3~\cite{Bechtle:2013wla} to make sure that the Higgs states other than $h$ satisfy the LEP, Tevatron, and LHC constraints. 

\section{Signal-to-background Analysis}

For our detector analysis, in order to directly reconstruct the \hobs\ (implying here a 125\gev\ Higgs state) we consider only the $\hobs\to b\bar{b}$ decay, since both the $b$-quarks are observable and also because in the models studied here the $h$ decays dominantly in this channel. The complete process analysed is then $pp\to (b)tH^\pm \to (b)b W^\mp W^\pm \hobs \to (b)bbb jj \ell\nu_\ell$, where one of the $W$-bosons (from either $H^\pm$ or $t$-quark) decays leptonically and the other hadronically.  The presence of a single lepton allows us to avoid multi-jet backgrounds, while requiring one hadronic $W$ avoids additional unseen neutrinos. The main background for this process is $t\bar{t}b(\bar{b})$, where either an additional $b$-tagged jet combines with a $b$-jet from a top decay or an additional $b\bar{b}$ pair mimics an $\hobs\to b\bar{b}$ decay.

In order to estimate the sensitivity obtainable at the $14\tev$ LHC, we generated the $t(b)H^-$ signal using Pythia 6.4.28~\cite{Sjostrand:2006za} with the MATCHIG~\cite{Alwall:2004xw} add-on to avoid double counting among the $gg\to tbH^-$ and  $gb\to t H^-$ processes, and all $t(b)WX,X\to b\bar{b}$ backgrounds with MadGraph5~\cite{Alwall:2011uj}. The parton showering and hadronisation for both the signal and the background was performed using Pythia 8~\cite{Sjostrand:2007gs} and the subsequent detector simulation with DELPHES 3~\cite{deFavereau:2013fsa}, using experimental parameters based on the ATLAS experiment. The $b$-tagging efficiency chosen was $\epsilon_\eta\tanh(0.03 p_T - 0.4)$, with $\epsilon_\eta = 0.7$ for central ($|\eta|\leq 1.2$, with $|\eta|$ being the rapidity) and $\epsilon_\eta = 0.6$ for forward ($1.2\leq|\eta|\leq 2.5$) jets, and the transverse momentum, $p_T$, in GeV.

To reconstruct our signal events and separate them from the background, we then proceeded as follows:
\begin{enumerate}[noitemsep]
\item Only events with $\geq 3$ $b$-tagged jets, $\geq 2$ light jets, one lepton ($e$ or $\mu$) and missing energy (\met) $\ge 20\gev$ were selected. For all objects we required $p_T \ge 20$\gev, $|\eta|\leq 2.5$ and $\Delta R$, their separation from other objects, $\ge 0.4$.
\item First the pair of light jets with an invariant mass, $m_{jj}$, closest to the $W$ boson mass, $m_W$, was chosen. An event was rejected if no pair satisfied $|m_{jj}-m_W|\leq 30\gev$.
\item The observed lepton was used to find the longitudinal component of the neutrino momentum, $p_{\nu,z}$, by imposing the mass constraint $m_{\ell\nu}=m_W$, thus attributing all $\met$ to a neutrino from a $W$ decay. Such solutions have a twofold ambiguity as a result of the quadratic nature of the constraint. Both solutions were kept when they were real. For complex solutions, a single real $p_{\nu,z}$ was retained, discarding the imaginary component.
\item The background can mimic the signal by combining a $b$-jet from a top decay with an additional $b$-tagged jet to reconstruct the $h$. Thus as a `top veto' was applied at this stage, but only in the large $m_{\hpm}$ region. This implies that an event was rejected if two $t$-quarks could be reconstructed from reconstructed $W$'s and any unassigned jets, with both satisfying $|m_{Wj}-m_t|\leq 20\gev$. The jets used may or may not be $b$-tagged. For $m_{\hpm}$ not too far above the $\hpm\to W^\pm \hobs$ threshold, one of the resulting $b$-jets combines with the $W^\pm$ to give an invariant mass $m_{bW}\approx m_t$ in a large fraction of the available phase space. Such signal events are cut if the top veto is applied at this point, negating the benefits of the background reduction. Therefore, for $m_{\hpm} \lesssim 350$\gev, the top veto was postponed until after the next step.
\item The pair of $b$-tagged jets with invariant mass, $m_{bb}$, closest to $m_{\hobs}\sim 125\gev$ was used to reconstruct the \hobs. The event was rejected if no pair satisfied $|m_{bb}-m_{\hobs}|\leq 15\gev$.
\item In the low $m_{\hpm}$ region, the top veto was applied at this stage in the same way as above, but excluding the $b$-jets used in $\hobs$ reconstruction. 
\item From the reconstructed $W$'s and remaining $b$-tagged jets, the best top quark candidate was identified as the one for which the $Wb$ combination had an invariant mass, $m_{Wb}$, closest to $m_t$. If the selected combination included one leptonic $W$ solution, the other was discarded. The event was rejected if there was no good candidate with $|m_{Wb}-m_t|\leq 30\gev$.
\item The reconstructed $\hobs$ was combined with the remaining $W$ to yield the discriminating variable $m_{W\hobs}$. If there were two leptonic $W$'s remaining, both values of $m_{W\hobs}$ were retained. For each $\mhpm$ considered, we placed a cut on the range of reconstructed $m_{W\hobs}$ which maximises the statistical significance, $S/\sqrt{B}$.
\end{enumerate}

\section{Results and discussion}

\begin{figure}[ht!]
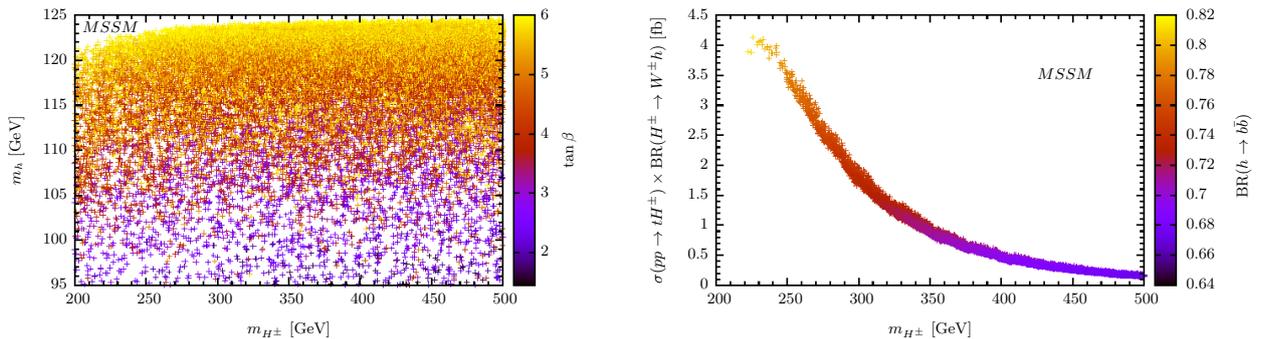

\vspace*{-0.35cm}
\centering
	\resizebox{0.49\textwidth}{!}{\input{./MSSM_mh_mhpm_tanb.tex}}
	\resizebox{0.49\textwidth}{!}{\input{./MSSM_mhpm_XS_mh.tex}}
\vspace*{-0.5cm}
\caption{$m_h$ as a function of $\mhpm$ and $\tan\beta$ in the MSSM (left) and $\sigma(pp\rightarrow t\hpm)\times
  \text{BR}(\hpm\rightarrow W^\pm h)$ as a function of $\mhpm$, with the heat map showing the BR($h \rightarrow b\bar{b}$) (right).}
\label{fig:MSSM}
\end{figure}

In the MSSM, for the selected \mhpm\ range, $m_h$ lying above 123\gev\ can only be reached (for $\mathcal{O}(1\tev)$ scanned values of the soft SUSY parameters) near the allowed upper limit of $\tan\beta$, as shown in the left panel of Fig.~\ref{fig:MSSM}. The right panel shows that such values of $\tan\beta$ and $m_h$ always yield a very poor $\sigma(pp\rightarrow tH^\pm)\times BR(H^\pm\rightarrow W^\pm h)$. Multiplying it with BR$(h\rightarrow b\bar{b})$ gives the total signal strength for a given point, which is further reduced, as is evident from the heat map in the panel. We therefore do not test this model against the collider analysis.

\begin{figure}[ht!]
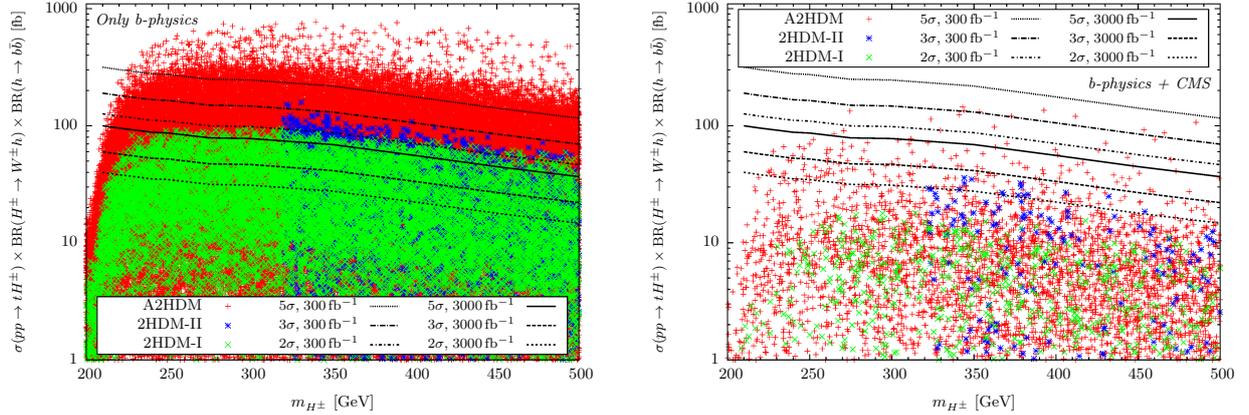

\centering
	\resizebox{0.49\textwidth}{!}{\input{./ALL2hdm_mhpm_XS.tex}}
	\resizebox{0.49\textwidth}{!}{\input{./CMS2hdm_mhpm_XS.tex}}
\caption{Signal strength for the points satisfying only the $b$-physics constraints (left) and additionally the CMS constraints on $\mu^{\gamma \gamma}$ and $\mu^{ZZ}$ (right) from the scans for the 2HDMs. Also shown are the contours for different statistical sensitivities expected for selected integrated luminosities at the LHC with $\sqrt{s}=14$\tev.}
\label{fig:results}
\end{figure}

For the 2HDMs, Fig.~\ref{fig:results} shows the results of the parameter scans along with the sensitivity expected from the collider analysis. For the $Z_2$-symmetric cases, we find a large number of points which are potentially discoverable at a high-luminosity (3000\,fb$^{-1}$) LHC, as seen in the left panel. The A2HDM shows even stronger signals, well within reach of even the standard luminosity (300\,fb$^{-1}$) LHC. In the right panel we consider the effect of imposing the signal strength, $\mu^X$, where $\mu^X=\sigma(pp\to\hobs\to X)/\sigma(pp\to \hsm\to X)$, with $\hsm$ being a $125\gev$ SM Higgs boson, measurements on $h$ for these points. For this purpose, the theoretical counterparts of $\mu^X$ (with \hobs\ replaced by $h$) were determined with HiggsSignals-v1.20~\cite{Bechtle:2013xfa} for $X= \gamma\gamma,\,ZZ$ and compared with the CMS measurements, $\mu^{\gamma \gamma} = 1.13 \pm 0.24$, $\mu^{ZZ} = 1.0 \pm 0.29$~\cite{CMS-PAS-HIG-14-009}. We note that, owing to significant deviations of $h$ from \hsm-like properties for most of the good points from the scans, those surviving after imposing the $\mu^X$ constraints have a significantly lower detection potential, particularly in the case of $Z_2$-symmetric 2HDMs. In the A2HDM, some points still remain testable even with 300\,fb$^{-1}$ luminosity. We therefore conclude that the $\hpm\to W^\pm h$ channel can be a useful probe of \hpm\ in the 2HDMs with high luminosities at the LHC.

\begin{acknowledgments}
This work was in part funded by the Swedish Research Council under
contracts 2007-4071 and 621-2011-5107. The work of S.~Moretti has been funded in part through the NExT Institute. The computational work was in part carried out on
resources provided by the Swedish National Infrastructure
for Computing (SNIC) at Uppsala Multidisciplinary Center
for Advanced Computational Science (UPPMAX) under Projects p2013257 and SNIC 2014/1-5.
\end{acknowledgments}

\bigskip

\bibliography{HPNP2015_Moretti}

\end{document}